\documentclass{ws-procs9x6-cpt16}
\begin{document}

\newcommand{\refeq}[1]{(\ref{#1})}
\def\etal {{\it et al.}}

\title{The NNbar Experiment at the European Spallation Source}

\author{M.J.\ Frost}

\address{Department of Physics and Astronomy, 
The University of Tennessee-Knoxville\\
Knoxville, TN 37996, USA}

\author{On behalf of the NNbar Collaboration}

\begin{abstract}
The observation of neutron to antineutron oscillation would be the first experimental evidence to show that baryon number is not a conserved quantity. It also provides an answer to the hypothesized post-sphaleron baryogenesis mechanism shortly after the Big Bang. The free oscillation time $\tau_{n \rightarrow \bar{n}}$ has a lower limit at $8.7\times 10^{7}$ seconds determined at ILL in 1994. Current beyond Standard Model theories of this oscillation time estimate the value to be on the order of $10^{10}$ seconds. A new experiment is proposed at the European Spallation Source that has 1000 times the sensitivity 
of the previous experiment, and would confirm the viability of those beyond Standard Model theories.
\end{abstract}

\bodymatter

\section{Motivation}
Modern cosmological models propose baryon number violation (BNV) as an explanation for the observed imbalance of matter and antimatter in the universe. BNV $(\Delta B \neq 0)$ decays can exist while maintaining $\Delta(B-L)=0$, a requirement of the Standard Model. Theorized BNV mechanisms such as proton decay $(\Delta B=1)$ occur at much higher energies (grand-unified theory), and any asymmetry arising from this would be erased during the electroweak sphaleron phase transition.\cite{electbnn} Thus, $\Delta(B-L)=0$ alone would not be a suitable basis for explanation of baryogenesis. $\Delta B=2$ processes like $n \rightarrow \bar{n}$ transitions probe energies above the Large Hadron Collider, but still below the phase transition energy, and would confirm the need to probe further at these scales or rule out post-spheralon baryogenesis.\cite{psb} Majorana particles\cite{majorana} provide a usable phenomenology for development of an experiment to observe BNV.\cite{nnbarstat}

Two-level, time-dependent systems oscillate between two states based on
the off-diagonal mixing term $\delta m$. This term distinguishes the two states from one another, and provides insight to their suppression modes, in this case
when an external magnetic field is present. This assumes, as required by CPT symmetry, that the neutron and antineutron masses are equivalent and that their dipole moments are equal and opposite. In a short time limit and low magnetic field configuration, the
probability of free transition of $n \rightarrow \bar{n}$ is
\begin{equation}
P(n(t)=\bar{n})\simeq[(\delta m)t]^2 = (t/\tau_{n \rightarrow \bar{n}})^2,
\label{nnbar:eq1}
\end{equation}
where $\tau_{n \rightarrow \bar{n}}$ is known as the oscillation time.

\section{The ESS and proposed NNbar experiment}

\begin{figure}
\includegraphics[width=\hsize]{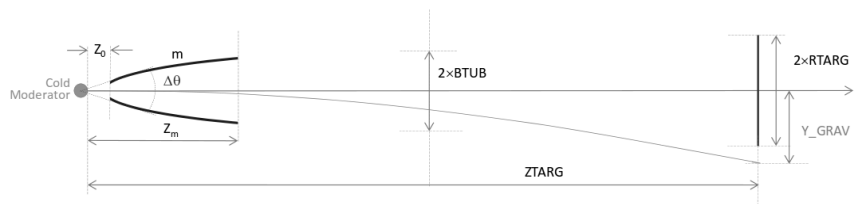}
\caption{A schematic view of the proposed NNbar experiment at ESS. The parameters specified are variable and subject to optimization.}
\label{nnbar:fig1}
\end{figure}

The European Spallation Source (ESS) is a pulsed spallation neutron source currently under construction in Lund, Sweden and will begin neutron production in 2019. The neutrons it produces are moderated and then transported to a suite of scattering instruments designed to characterize super-conducting, inorganic, engineering, and biological materials.\citep{essweb} The neutrons have a typical velocity of 1200 m/s, which is also well suited to fundamental neutron physics experiments such as NNbar. Using this intense source of cold neutrons, along with other advancements in thermal neutron optics, the experimental sensitivity of NNbar can be improved 1000 times or better than the last attempt at ILL\cite{illnnbar} during a 3-5 year running time.
The baseline configuration seen in Fig.\ \ref{nnbar:fig1} incorporates an ellipsoidal focusing reflector\cite{yuriellipse} designed to enhance the cold intensity at an annihilation target $\sim200$ meters away from the moderator.

\section{Novel optical concepts and sensitivity impact}
Simulations have shown that the ellipsoidal reflector concept will significantly increase the cold neutron transport from moderator to annihilation target. 
Beyond that, nested reflectors, lobed reflectors with multiple focal points, and nanoparticle quasispecular reflectors\cite{valera_nano} can provide further enhancement in both sensitivity and value.


\section{Conclusion}
The observation of neutron to antineutron oscillation would be the first experimental evidence to show that baryon number is not a conserved quantity; one of the three Sakharov conditions required for baryogenesis. A new experiment with 1000 times higher sensitivity can now be built given modern technological advancements in thermal neutron instrumentation at the ESS.

\section*{Acknowledgments}
Many thanks to Yuri Kamyshkov for subject guidance and Mike Snow for encouragement to attend the conference meeting. This research was funded by the University of Tennessee-Knoxville Office of Scientific Outreach and by the Department of Energy, High Energy Physics.

\end{document}